%% file: main.tex
\documentclass[sigconf]{acmart}
\usepackage{graphicx}
\usepackage{graphicx}
\usepackage{subcaption}
\usepackage{comment}
\usepackage{caption}
\usepackage{siunitx}
\usepackage{booktabs}
\usepackage{pifont}
\usepackage{hyphenat}
\usepackage{algorithm}
\usepackage{algorithmic}

\graphicspath{{./images/}}
\usepackage{pdfpages}

\setcopyright{acmcopyright}
\copyrightyear{2021}
\acmYear{2021}
\acmDOI{xx}

\acmConference[draft]{draft }{Oct 03--05, 2021}{Shanghai,China}
\acmBooktitle{draft}
\acmPrice{15.00}
\acmISBN{XXX-X-XXXX-XXXX-X/XX/XX}

\begin{document}

\title{NetDAM: Network Direct Attached Memory with Programmable In-Memory Computing ISA}

\author{Kevin Fang}
\affiliation{%
  \institution{Cisco Research \& Development Center}
  \city{Shanghai}
  \country{China}
}
\email{zhiyfang@cisco.com}

\author{David Peng}
\affiliation{%
	\institution{Cisco Research \& Development Center}
	\city{Shanghai}
	\country{China}
}
\email{yupeng@cisco.com}

\input{sections/abstract}

\vspace{1em}

\maketitle

\input{sections/introduction}

\input{sections/architecture}

\input{sections/allreduce}

\input{sections/conclusion}
\input{sections/acknowledgment}

\bibliographystyle{ACM-Reference-Format}
\bibliography{references.bib}

\end{document}

%% file: sections/abstract.tex
\begin{abstract}

Data-intensive applications like distributed AI-training  may require multi-terabytes memory capacity 
with multi-terabits bandwidth. We directly attach the memory to the ethernet controller with some programable logic to design an efficient hardware "template" for  \emph{Memory pooling} 
and \emph{in-memory / in-network computing}.

We built an FPGA prototype of the NetDAM, and we demonstrate MPI-Allreduce communication case, the NetDAM can be used as a software and hardware friendly programmable architeture with high performance alternative for RDMA. 

\end{abstract}


\ccsdesc[500]{Computer systems organization~Cloud computing}
\ccsdesc[500]{Networks~Transport protocols}
\ccsdesc[500]{Hardware~Networking hardware}

\keywords{Mempool, In-Memory Computing,In-Network Computing, Programmable ISA.}

%% file: sections/introduction.tex
\section{Introduction}
\label{sec:introduction}

In a conventional computer, the processing and memory units are physically separated. Consequently, significant data need to be moved during computation, which creates a performance bottleneck commonly called "von Neumann Bottleneck". Domain Specific ASICs(DSA) start to address this problem, however, ultra-high bandwidth and ultra-low latency communication is still required for DSA accelerator and CPU. 

\subsection{Intra-host vs Inter-host communication}

PCIe is widely used for intra-host communication, direct memory access (DMA) is used for inter-chip communication. RDMA~\cite{rdma} over Ethernet simply extend the DMA operation to inter-host network. The go-back-N method require lossless Ethernet, DCQCN ~\cite{DCQCN} is designed for congestion control and reliable data delivery. PFC based congestion control mechanism introduce significant latency, even more it may not work well at large scale and cloud based virtual private network environment, some vendors like fungible~\cite{fungible} use TCP Offload Engine on NIC with iWARP~\cite{iwarp}, HPCC~\cite{hpcc} NDP~\cite{ndp} are designed to eliminate the PFC but require special Ethernet switch to support In-network-telemetry(INT) or NDP. Simultaneously, ~\cite{pcieperf} shows that PCIe, alongside its interaction with the root complex and device drivers, can significantly impact the performance of end host networking. GenZ, CCIX, CXL are designed to address such problem.

Therefore,we re-examine the \emph{intra-host} and \emph{inter-host} communication protocols,Intra-host protocol often use \emph{shared memory mode}, the inter-host protocol often use \emph{message passing mode}, the design principal has significant difference:

\begin{itemize}

	\item\textbf{topology:} intra-host communication has dedicated topology which could easily use fixed algorithm(eg. DFS) to encoding devices address. message routing is very simple. inter-host communication are topology independent, multi-path and overlay transport may cause message routing more complicated.

	\item\textbf{latency:}The round-trip-time between intra-host chips only sub-nanoseconds, inter-host communication may much high and unpredictable during congestion via multi-path.

	\item\textbf{loss:} intra-host communication has dedicated topology with lossless circuits, meanwhile congestion or interim node failure may cause packet drop on inter-host communication

	\item\textbf{coherency:} The round-trip-time between intra-host chips only sub-nanoseconds, so coherency mechanism implementation is much easier and more efficient than inter-host communication. 

	\item\textbf{ordering:} Intra-host communications are strictly ordered to keep the cache coherency. Inter-host communication may cause Out-Of-Order under multi-path scenario.

	\item\textbf{flit size:} coherency and other real-time communication scenario require small flit size, many intra-host protocols(eg. CXL) are designed to use single flit just carry one cache-line(64B). The flit size in inter-host protocol(eg. Ethernet) is much larger(1500B).
\end{itemize}

 Directly extend \emph{intra-host} protocol to \emph{inter-host} (eg. RDMA)need carefully process the congestion control and coherence issue, however the host PCIe link and cache coherence processing  may introduce high latency and unpredictable jitters. congestion control module may use more buffer to absorb jitters but it may cause more delay.

 Directly extend \emph{inter-host} protocol to \emph{intra-host} (eg. NanoPU~\cite{nanopu}) require processor attached to the Ethernet controller, packet directly send and operate at register level which may loss programming flexibility and increase complexity when application need to use other sophisticated processors.
 
Therefore, "Dam" is required at the barrier of host to provide unified memory access to absorb the bursts and smoothly transition the flit size, "batch-mode" with large flit size for \emph{message passing interface} during inter-host communication, while small flit size to implement cache coherency for \emph{shared memory mode}.

\begin{figure}[h]
	\centering
	\includegraphics[scale=0.7]{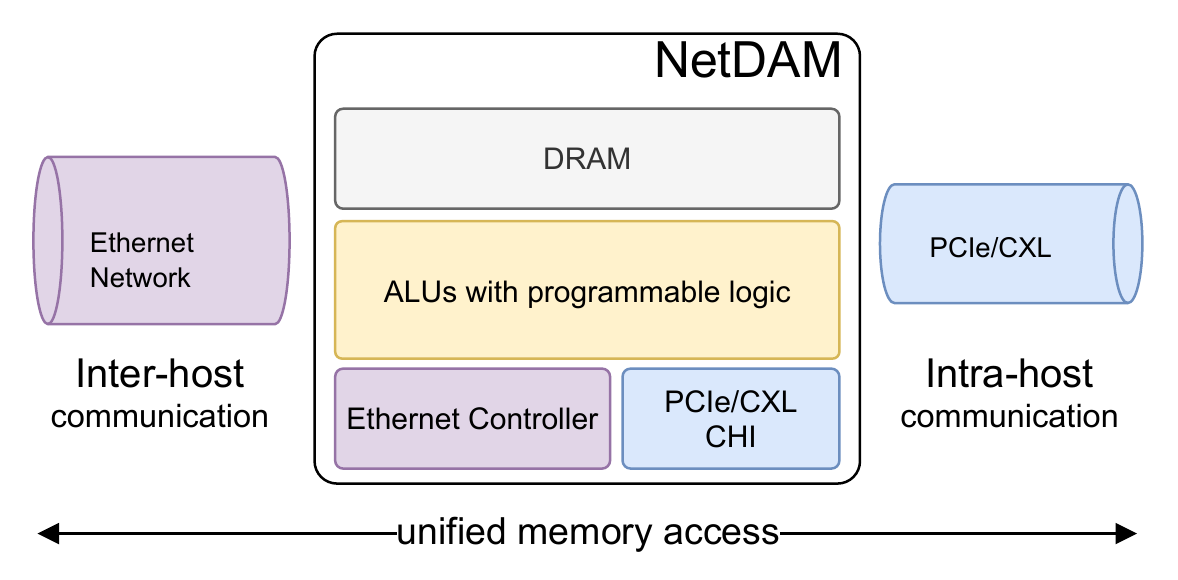}
	\caption{NetDAM function block}
	\label{Fig:netdam_arch}
\end{figure}

NetDAM is designed to bridge the \emph{intra-host} and \emph{inter-host} protocols by directly share memory with additional instruction level support for \emph{in-memory} and \emph{in-network} computing.  With this architecture, CPU / Domain Specific Accelerator / other storage component could directly attach to netDAM via AXI or CHI or PCIe/CXL bus and share the unified memory pool.

\begin{figure}[h]
	\centering
	\includegraphics[scale=0.55]{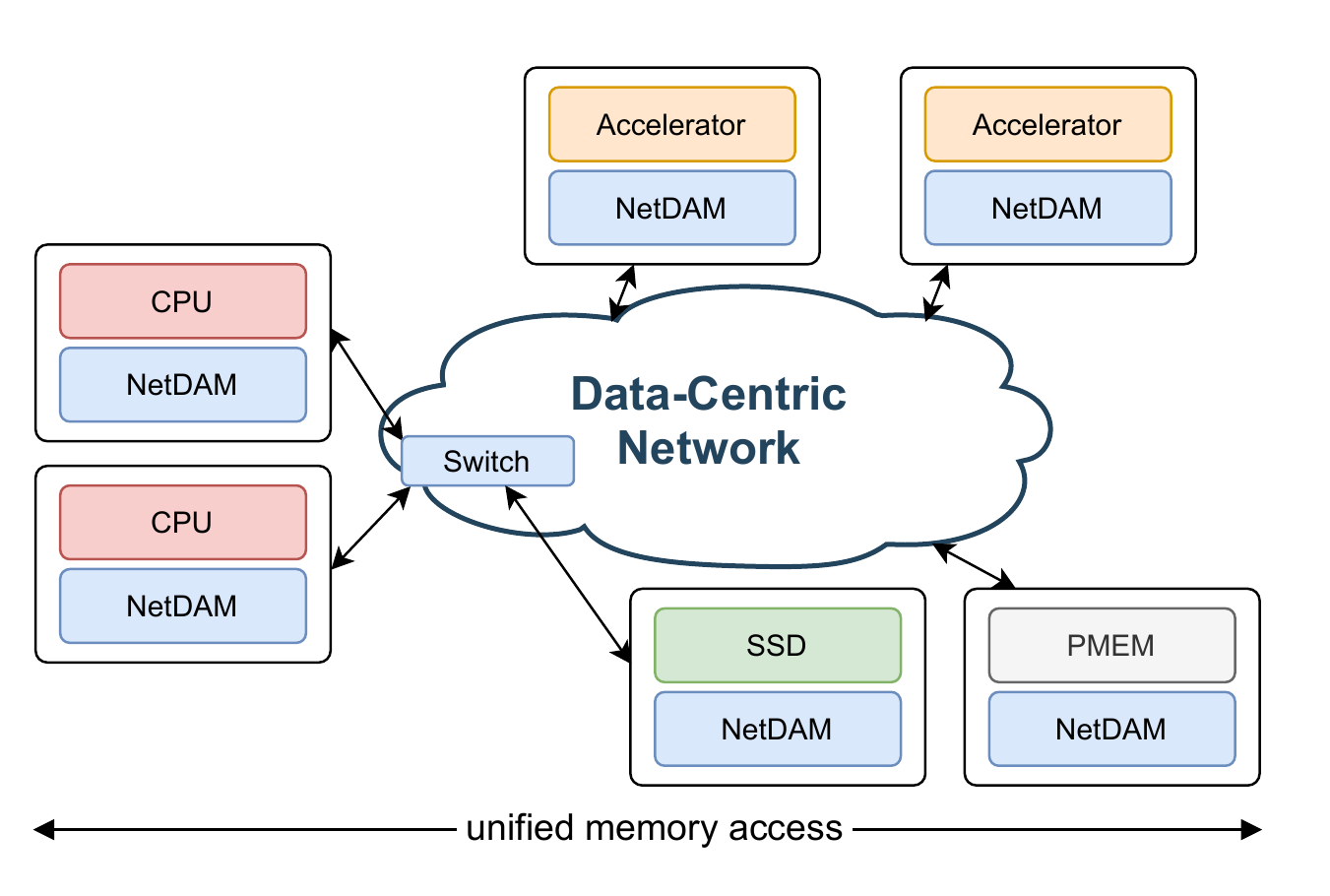}
	\caption{NetDAM interconnection system}
	\label{Fig:netdam_system}
\end{figure}

\noindent \textbf{Paper Organization:} We present the system architecture in Section~\ref{sec:architecture}. We demonstrate MPI-allreduce case by add reduce-scatter and all-gather instruction based on NetDAM programmable ISA in Section~\ref{sec:allreduce}.We conclude in Section~\ref{sec:conclusion}.

%% file: sections/architecture.tex
\section{System Architecture}
\label{sec:architecture}

In this section, we present the details about system architecture and design trade-off.

\subsection{Overview}

Direct memory access(DMA) at 200Gbps/400Gbps may highly impact the system memory, crypto,compression and regex offload may amplify this issue. Network I/O and system memory segregation by programmable Memory on NIC is important to address DMA congestion issue and DPU offloading issue. 
NetDAM is designed to provide programmable memory access on both side. Remote directly memory access to NetDAM region over Ethernet may get deterministic latency, thus simplify the congestion control mechanism. In-host side, it could simply use memory interface(memif) and related poll mode driver to get packet or even special header to mitigate the system memory bandwidth usage.

\subsection{NetDAM packet format}

NetDAM is a packet based protocol which combines the instructions and data. 

\begin{figure}[h]
	\centering
	\includegraphics[scale=0.75]{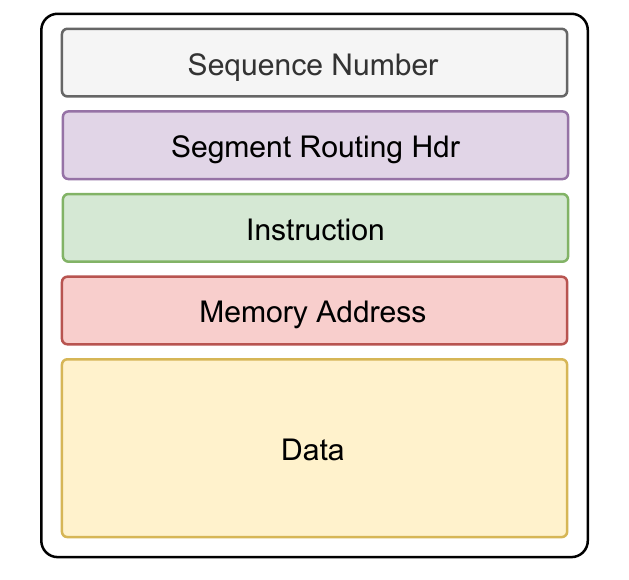}
	\caption{NetDAM packet format}
	\label{Fig:netdam_system}
\end{figure}

\begin{itemize}
	
	\item\textbf{Sequence:} is used for packet ordering and reliable transmit.
	
	\item\textbf{Segment Routing Header:} is used for leverage multi-path to avoid datacenter network congestion. Meanwhile, it could be used for dataflow computation. Many distributed systems use Directed acyclic graph(DAG) to abstract the computation job, Segment Routing Header could be a chaining function to processing packet on different node. Detailed information could be found in subsection ~\ref{subsec:transport_layer}.
	
	\item\textbf{Instruction:} NetDAM provide basic \emph{READ},\emph{WRITE},\emph{CAS} and \emph{MEMCOPY} instructions which operate in SIMD mode. NetDAM support programmable ISA extension, we reserve multiple bits in this field, user could define their own instructions for different computation jobs, detailed information could be found in subsection ~\ref{subsec:isa}.
	
	\item\textbf{Address:} The instruction operate data memory address. IOMMU may implement on NetDAM for Virtual Address and Physical Address translation. Remote Memory could also mapping to local Virtual Address by this IOMMU, detailed information could be found in subsection~\ref{subsec:sys_address}.
	
	\item\textbf{Data:} has variable length based on instruction definition, many instruction could be operated in SIMD mode, especially for inter-host communication. Data length could be 9000B which means SIMD could leverage multiple ALUs on NetDAM to operate $\approx$ 2048 x float32 in parallel.
	
\end{itemize}

\subsection{Transport Layer}
\label{subsec:transport_layer}

NetDAM design principal is providing ultra-low latency and ultra-high bandwidth transport with necessary hardware dependencies, IP/UDP over Ethernet is used to carry NetDAM packet for inter-host communication.

\textbf{Deterministic Latency:} NetDAM has fixed pipeline to processing packet by eliminate PCIe DMA and bypass snoop for cache coherency. Packet could be ACK with deterministic latency, we test wire-to-wire SIMD read 32 x float32 data from DRAM, average latency is $618$ nanoseconds, jitter is $39$ nanoseconds, max latency is only $920$ nanoseconds, which is much faster than RoCE. Latency is the key indicator for congestion control (eg. SWIFT~\cite{swift}), We need to emphasize isolate the intra-host DMA to implement inter-host deterministic latency communication is very useful to simplify the congestion control algorithm. 

%FPGA  mean 763 std 46 min 728  median 752 max 1088
%switch mean 145 std  7 min 136  median 144 max 168 

\textbf{Reliable Transmit:} is optional. One reason is lossless Ethernet with virtualization or container overlay support will introduce significant overhead. Another reason is many distribute applications could design \emph{idempotent interface}, simply re-transmit does not impact the result, \emph{atomic instruction} could be added to implement \emph{idempotent operators}.

\textbf{Relax Order:} Commutative operations allows Out-Of-Order execution, especially we have memory address field in each packet to isolate operating memory space. However we provide sequence field in the packet, user could add optional reorder module in programming logic for ordering execution.

\textbf{Multi-Path:} Many datacenter network topology use fat-tree while some HPC cluster use 2D-Torus 3D-Torus. NetDAM design Segment Routing Header in UDP(SROU ~\cite{ruta} ) enable topology independent transport, source node could select dedicated path to avoid switch buffer overrun and fully utilize the fabric bandwidth. function callback could add in segment routing stack for chaining computations over multiple node, we will show a ring-allreduce demo in section\ref{sec:allreduce}.

\subsection{Programmable ISA} 
\label{subsec:isa}
NetDAM instruction is more like a RPC(Remote Procedure Call) for software friendly development, it has dedicated memory space for Request and Complete Command Queue pairs. software could simply write the NetDAM packet to Request Queue memory address, and fetch from Complete Queue. For the inter-host communication case, software could simply use UDP socket send NetDAM packet to NetDAM device.

NetDAM "template" only define some basic memory access instructions(WRITE, READ, MEMCOPY, Atomic) and leaves multiple bits in instruction field for user defined behavior. 

For the neural network case, user may define SIMD(ADD, SUB, MUL, XOR, MIN, MAX) and compute them directly near the memory, on-chip ALUs will be used for accelerate computation, HBM-PIM~\cite{hbm_pim} could be attached to NetDAM for in-memory acceleration in the future.

This architecture also allows user defined your own circuit logic to build DSA IPCore and directly connect to NetDAM via AXI bus for adaptive computing. For MPI all-reduce communication case, Reduce-Scatter and All-Gather instruction could be added. For DPU offload case, compress, crypto, hash and longest prefix match instruction could be added.

\subsection{Memory Addressing}
\label{subsec:sys_address}

Each NetDAM device has its own memory address space and mapping to intra-host network, additional IOMMU could be added for virtualization support and mapping remote memory to local host.

A special address pool could be used for NetDAM pkt Request Queue and Complete Queue. 

NetDAM directed attached DRAM memory address could be mapping to host user-space(also support the VM/Container deployment), Shared Memory Packet Interface (memif)~\cite{memif} could be implement on NetDAM for general Ethernet packet.

\begin{figure}[h]
	\centering
	\includegraphics[scale=0.95]{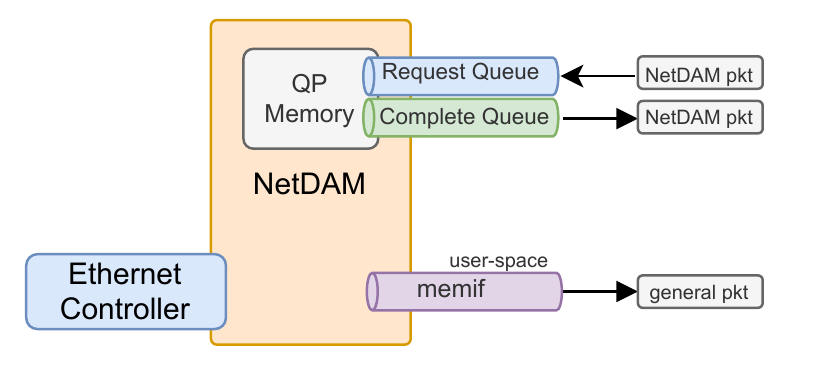}
	\caption{NetDAM memory interface}
	\label{Fig:memory_allocation}
\end{figure}

NetDAM device could be operated in standalone mode and directly attached to switch, multiple NetDAM device with switch construct a big memory pool with multi-terabytes memory capacity with multi-terabits bandwidth.

\begin{figure}[h]
	\centering
	\includegraphics[scale=0.85]{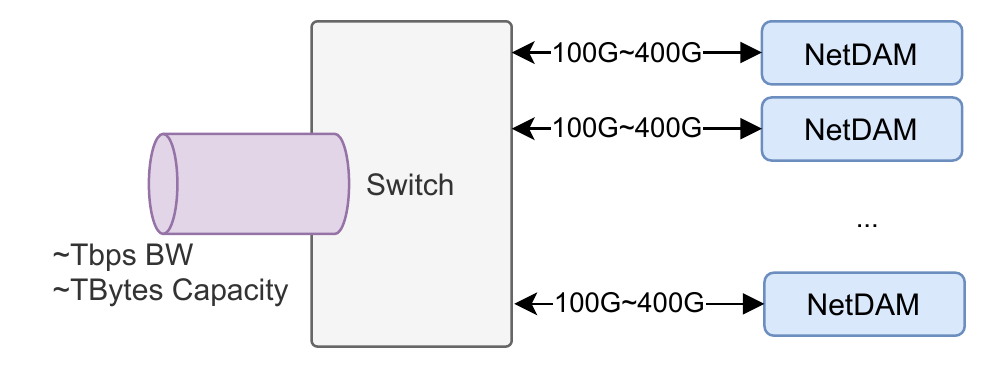}
	\caption{NetDAM memory pool}
	\label{Fig:mem_pool}
\end{figure}

Each NetDAM could implement a local IOMMU to translate Global Virtual Address to NetDAM device IP address with NetDAM Local Address. Such IOMMU function could be implement on programmable switch, Virtual IP address cloud be configured on switch and provide address translation to NetDAM device.

\textbf{Incast Avoidance} The global memory pool could be operated in block interleaved mode, Thus many-to-one communication could be equally load balance to multiple NetDAM device, the receiving host could pull them back from global memory pool based sequencing and rate-limited READ command, the incast problem can be easily avoid without complex congestion control mechanism.

\subsection{Security }
\label{subsec:security}
Security problem is a concern when sharing memory and executing instruction remotely, SDN controller could act as a MMU to simply apply malloc/free request and translate request to access-control-list and apply to each NetDAM or in datacenter switch. \emph{encryption-write}  and \emph{decryption-read} instruction could be added for secure computing.

%% file: sections/allreduce.tex
\section{MPI Allreduce}
\label{sec:allreduce}
MPI Allreduce communication is the key bottleneck in Distributed AI training platform. We implement MPI-Allreduce instruction on NetDAM and demonstrate it could be much faster than RoCE.

\noindent\textbf{Ring-Allreduce}\cite{ringallreduce} is a high efficient all-reduce algorithm designed by Baidu research, then widely used in Horovod library. The collective communication is based on ring topology which may fully utilize the network bandwidth and mitigate congestion.

We implement \textbf{Ring Reduce-Scatter} and  \textbf{Ring All-gather} instruction on NetDAM, The prototype use 2 x Xilinx Alveo U55N FPGA with build-in HBM, Each blade implement 2 independent NetDAM device. Each NetDAM device has one 100G Ethernet Port with 2GB HBM memory. All 4 NetDAM device connect to a Cisco Nexus 93180FX switch.

\subsection{Ring Reduce-Scatter}

Ring Reduce-Scatter is a chaining function to add the parameter node by node. Node1 send \emph{A1} to Node2, Node2 calculate \emph{A1+B1} then send it to Node3,  Node3 send \emph{A1+B1+C1} to Node4. Node4 write \emph{A1 + B1 + C1 + D1} in local memory.

\begin{figure}[h]
	\centering
	\includegraphics[scale=0.85]{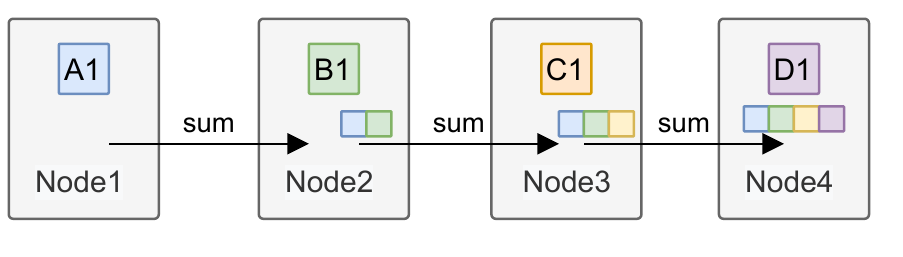}
	\caption{Ring Reduce-Scatter}
	\label{Fig:rrs}
\end{figure}

RoCEv2 based implementation require multiple DMA and multiple load/store on CPU, the temporary sum A1 + B1 may require separated memory space to store to avoid side-effect. The sum operation need explicit instruction in each iterations, an explicit synchronization barrier required between iterations which may introduce more latency.

\begin{figure}[h]
	\centering
	\includegraphics[scale=0.95]{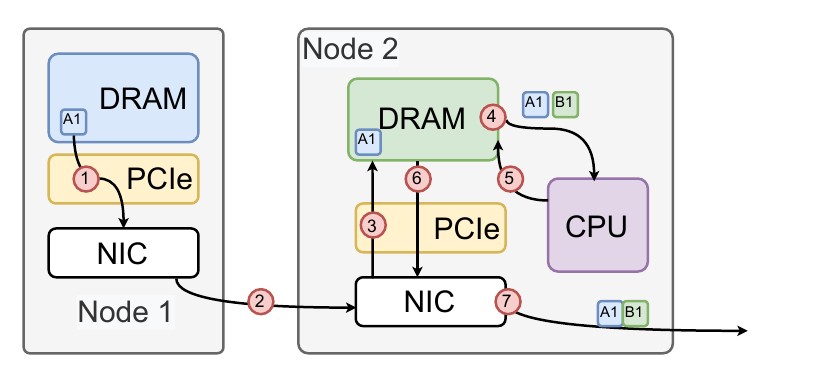}
	\caption{RoCE Reduce-Scatter}
	\label{Fig:rdma_reduce}
\end{figure}

NetDAM implementation is much faster than RoCE, A1 directly fetch from DRAM on Node1 and send to Node2 packet buffer SRAM, Node 2 execute the instruction in ALU to add B1 and store the sum result in SRAM, then based on Segment Routing header in SRAM to self-route the packet to Node 3. Traditional CPU may only has AVX512 instruction support, each cycle may only support 32x float32 value add operation. NetDAM could leverage directly memory access and implement multiple ALUs to support 2048 x float32 add operation with single instruction.

\begin{figure}[h]
	\centering
	\includegraphics[scale=0.95]{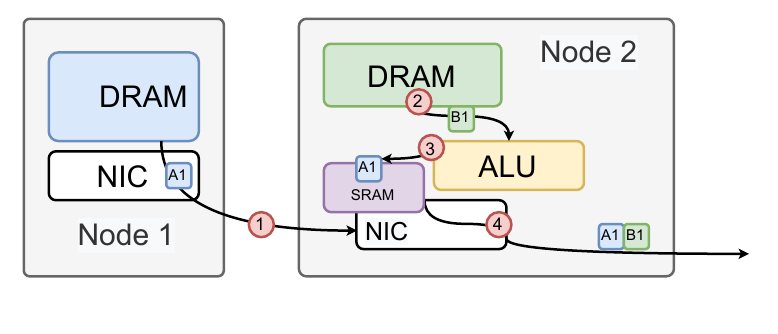}
	\caption{NetDAM Reduce-Scatter}
	\label{Fig:ndam_reduce}
\end{figure}

\textbf{Idempotent} is important for error handling in distributed system, the reduce-scatter operator is idempotent in interim node(Node2 and Node3), because all computation is based on packet buffer modification, no side-effect on local memory, but the last node need to write to local memory which is not idempotent. we defined a block based hash algorithm to keep the last hop idempotent. \emph{block-hash} instruction added to calculate block-hash, each blocks may contains 2048 x float32 data. The \emph{Reduce-Scatter} could carry this hash value, when last hop local memory hash is same with the instruction, NetDAM write the packet data to local memory else drop the packet.

\subsection{Ring All-Gather}

Ring All-Gather is simpler than Reduce-Scatter, it only require write the packet payload to each NetDAM device local memory, then use segment routing header route to next node, finally an ACK generated to send to the controller.
Multicast/Broadcast may implement in the future to accelerate the all-gather procedure.

\subsection{Evaluation}

We build a 4xNodes RoCEv2 Platform to compare with 4xNodes NetDAM. RoCEv2 platform use Mellnanox CX516A NIC, 2x Intel Xeon Gold 6230R CPU, 12x DDR4 32G@2933Mhz DRAM. Mellanox HPC-X and OFED installed.

We execute 536,870,912 x float32 allreduce, the native MPI Allreduce takes 2.8 seconds, the ring-based allreduce use 2.1 seconds. NetDAM will only takes xxx microseconds(initial testing result is 400ms, we are fixing some bug, it might be even faster).

%DataSize(Bytes)         Num of float       Native Allreduce         Ring Allreduce
       %128                   32                258.3us                 91.5us
       %512                  128                 11.5us                 24.3us
      %2048                  512                 18.4us                 26.8us
      %8192                 2048                 30.6us                 37.4us
     %32768                 8192                 63.0us                 73.1us
    %131072                32768                124.9us                171.9us
    %524288               131072                426.7us                553.4us
   %2097152               524288               1397.1us               2054.5us
   %8388608              2097152               5818.5us               7968.5us
  %33554432              8388608              23183.7us              31015.2us
 %134217728             33554432             136283.2us             129064.3us
 %536870912            134217728             471419.2us             524395.7us
%2147483648            536870912            2798183.4us            2109793.8us

%% file: sections/conclusion.tex
\section{conclusion}
\label{sec:conclusion}

We re-examine the intra-host communication bus(eg, PCIe, CXL, AXI, CHI..) and inter-host Ethernet network, there are significant differences in topology, latency, loss tolerance, coherency, ordering and flit size. We conclude that directly use one protocol for both communication network or use overlay technology does not efficient. 

Based on our decade experience in Cisco Quantum Flow Processor~\cite{qfp}, and from software/hardware programmable friendly perspective, we conclude directly attach memory to Ethernet controller and provide programmable instruction level memory access is the better way to isolate the intra-host and inter-host network, it could be smoothly transition the flit size, "batch-mode" with large flit size for \emph{message passing interface} during inter-host communication, while small flit size to implement cache coherency for \emph{shared memory mode}. 

We also leverage the Segment Routing concept to support multi-path loadsharing in datacenter network to reduce congestion, and the global memory pool with block interleaved addressing can be used to avoid incast problem. 

We build a prototype on FPGA to verify this idea, memory access latency and jitter is much lower than RoCE. We demonstrate the programmable ISA based on MPI Allreduce case which is critical for hyper-scale AI training fabric, by simply adding few domain specific instruction, it shows incredible performance improvement than RoCE, we will add dynamic sparse matrix compression and implement in-memory optimizer in the future.

In the future, once we get the CXL based FPGA and CPU, we will implement high speed intra-host communication stack (eg. memif). 

NetDAM architecture is not only for computation and memory acceleration, but also could accelerate storage network including NVMeOF and Persistent Memory support. Even more, cloud service provider(CSP) can use this infrastructure for serverless computing. We hope our results will encourage other researchers to push these ideas further.

%% file: sections/acknowledgment.tex
\section{acknowledgment}
\label{sec:acknowledgment}

Special thanks Xilinx provide demo equipment and technical support for this research project.